\documentclass{article}

\usepackage{times}
\usepackage{uist}

\usepackage{url}
\usepackage{hyperref}
\usepackage{times}
\usepackage{rotating}
\usepackage{colortbl}
\usepackage{epsfig}
\usepackage{alltt}
\usepackage{xspace}
\usepackage{graphics}
\usepackage{graphicx}
\usepackage{amssymb}
\usepackage{amsmath}
\usepackage{amsfonts}
\usepackage{ulem}
\usepackage{multirow}
\usepackage{tikz}

\definecolor{LightGray}{gray}{0.8}

\begin{document}



\title{An HCI View of Configuration Problems}


\author{
Tianyin Xu, Vineet Pandey, and Scott Klemmer\\
University of California, San Diego\\
\{tixu, vipandey, srk\}@eng.ucsd.edu
}

\maketitle




\tolerance=400

\section{ABSTRACT}

In recent years, configuration problems have drawn tremendous attention
because of their increasing prevalence and their big impact on system availability.
We believe that many of these problems are attributable to today's configuration interfaces
that have not evolved to accommodate the enormous shift of the system administrator group.
Plain text files, as the {\it de facto} configuration interfaces, assume administrators' understanding
of the system under configuration. They ask administrators to directly edit the
corresponding entries with little guidance or assistance.
However, this assumption no longer holds for today's administrator group which has
expanded greatly to include non- and semi-professional administrators.
In this paper, we provide an HCI view of today's configuration problems, and
articulate system configuration as a new HCI problem.
Moreover, we present the top obstacles to correctly and efficiently configuring software systems,
and most importantly their implications on the design and implementation of new-generation configuration interfaces.

\section{1. INTRODUCTION}



System configuration is a typical human-computer interaction (HCI) process.
The human administrator interacts with the computer system by setting configuration
parameters, in order to control the system's runtime behavior.
During the interaction, the administrator might encounter situations that
the system fails to work as expected, referred to as {\it configuration
problems}. The incorrect configuration settings are called {\it misconfigurations}.
In recent years, configuration problems have drawn tremendous attention
because of their increasing prevalence and the severity of misconfigurations.
For example, our study shows that configuration problems account
for 27\% of technical support cases in a major storage company in US~\cite{yin:11}.
A recent study on Hadoop systems reports that misconfigurations are the dominant causes of failures
in terms of both customer cases and support time~\cite{rabkin:13}.

Different from end-user applications, misconfigurations of systems may have a big impact.
For example, a recent misconfiguration at Microsoft left Azure service unavailable to their western European
customers for more than 2 hours~\cite{azure_outage}.
In 2011, a misconfiguration of Amazon's EC2 service caused the cloud crash,
affecting numbers of Internet services running on top of it~\cite{amazon_cloud_crash}.
In 2009, a misconfiguration brought down the entire ``.se'' domain for more than an hour, affecting
almost 1 million end hosts~\cite{se_domain}.

\begin{table*}[!t]
\small
\centering
\begin{tabular}{|p{3.5in}|p{3.2in}|}
\hline
\multicolumn{1}{>{\columncolor{LightGray}}c|}{\textbf{Characteristics of Configuration Problems (Section 2, 3)}} &  \multicolumn{1}{|>{\columncolor{LightGray}}c}{\textbf{Design Implications}} \\ \hline

{\bf (1) The shift of administrators.} The system administrator group has expanded greatly
to include non- and semi-professional administrators.
&  New interface for system configuration is desired to accommodate the shift of the administrator group. \\
\hline

{\bf (2) Configuration and Programming are anti-correlated.} 
Administrators and programmers (including scripting) form
different communities and have different skill sets.            
& System configuration should be studied as a separate problem from programming. The principles of building (end-user)
programming interface might not be applicable to configuration. \\
\hline

{\bf (3) The separation of understanding and manipulation.} 
The separation of user manuals and configuration
files causes administrators' cognitive difficulties and errors.

\vspace{0.2pt}
Google search is the first choice for users to solve
configuration problems than user manuals. 
& 
System vendors should not assume that manuals can help users solve their configuration problems. Instead, the 
configuration interfaces should integrate the information in user manuals.\\
\hline

{\bf (4) Difficulties rather than errors.} In more than 65\% cases, the configuration problems are 
administrators' difficulties (e.g., finding related parameters, setting values)
rather than committing errors. & The configuration interface should try to guide and educate administrators rather than directly asking
them to input values of configuration parameters.\\
\hline
\hline

\hline
\multicolumn{1}{>{\columncolor{LightGray}}c|}{\textbf{Cognitive Obstacles
to Configuring Systems (Section 4)}} &  \multicolumn{1}{|>{\columncolor{LightGray}}c}{\textbf{Design Implications}} \\ \hline

\vspace{0.5pt}
{\bf (1) Lack of guidance and information.} This is the major cause of today's administrators' configuration problems. 

\vspace{4pt}
This results in two challenges towards
configuration: 1) Finding the right parameters relevant to tasks from the entire parameter set; 
and 2) Setting the parameters' values to achieve the intended system behavior.
&

Configuration interfaces should be more
informative to help administrators address the two challenges.

\vspace{3pt}
To address the first challenge, configuration interfaces should provide administrators with
dependency, correlation, and association information regarding to their settings.

\vspace{3pt}
To address the second challenges, constraints, potential impact, and working examples should
be provided by the interfaces. \\

\hline

{\bf (2) Inconsistency and ambiguity.} Inconsistency of interface appearance,
correctness rules, and system behavior are one major cause of configuration problems,
including both difficulties and errors.   & Conceptually integrity should be carefully maintained for configuration interfaces,
                                            between interfaces and user manuals, and between interfaces and system behavior. \\
\hline

{\bf (3) System and control complexity.} A significant portion of users' configuration difficulties are caused by their incapability
in dealing with system and control complexity.

\vspace{0.5pt}
Our hunch is that non-professional administrators have less performance and security concerns as professional administrators.
&  We should decouple the configuration interface for ``dummies'' and professional administrators, in a similar way as~\cite{coar:98, kabir:99}.

\vspace{5pt}
Configuration parameters with different necessity and skill prerequisite should be separated in the different interfaces. \\
\hline


{\bf (4) Lack of environment awareness.} This is one common difficulty of diagnosing and resolving configuration-related system anomalies. 
 The environment of a running system includes its underlying stacks (e.g., OS)
 and co-running software. &

Configuration interfaces should help administrators recognize the environment information correlated to the configuration
settings, for example, constraints, entities, and resources.

\\

\hline

{\bf (5) Lack of technical support.} Administrators have difficulties in using Internet as technical support.
Many questions of configuration problems are not answered or with unsatisfied answers. &

Internet-based technical support services should try to reduce the response time and improve the efficiency
of diagnosing/solving configuration problems.
\\
\hline
\end{tabular}
\vspace{3pt}
\caption{\bf Our findings on configuration problems and their implications for configuration interface design}
\vspace{2pt}
\label{tab:summary}
\end{table*}

The prevalence of configuration problems comes with the proliferation
of free, open-source system software, as well as the boost of
economical computing utilities. The cost of deploying systems to provide services
keeps decreasing and is affordable to small business and even end users.
For example, the cost of running an Internet application on Amazon's public cloud today
is 10 times lower than a decade ago~\cite{andreessen:11}. Nowadays, there are
millions of users providing services on Amazon EC2 using open-source software such as LAMP
for web sites, Hadoop for big data analysis, and OpenStack for computing cloud.
Accordingly, the {\it system administrator} group has significantly
been expanded by semi- and non-professional administrators who
have limited technical expertise and interest in system management.

Unfortunately, today's configuration interfaces fail to accommmodate this
shift of the administrator group. 
The {\it de facto} configuration interfaces are still plain text files.
Despite great accessibility and scalability, file-based interfaces assume that
administrators have good understanding of the configuration knobs and their
impact on system behavior.
Administrators are directly asked to edit the corresponding entries in the file with little guidance or assistance.
However, this assumption no longer holds for today's administrators.
In our study (c.f., Section 3.1), we find that a significant portion of administrators have difficulties in finding the right
configuration knobs and in setting these knobs.

We advocate the HCI community to take the responsibility of
improving the configuration interfaces to help system administrators
configure systems correctly and efficiently.
In essence, configuration problems are derived from
administrators' cognitive difficulties and errors when they interact
with the configuration interfaces. Improving the interfaces has a great potential to help
administrators avoid cognitive biases. Thus, it attacks the
root causes of many configuration problems. In addition, defending against
misconfigurations at interface level is more time- and cost-efficient
than dealing with the resulting system failures and anomalies.
However, configuration interfaces have been overlooked in the past decades.
To this day, We possess little understanding of administrators' configuration problems,
including the difficulties encountered by them and the errors made by them.
Consequently, we do not know how to design ``good'' configuration interfaces
to help them.



We would like to note that the nature of system configuration distinguishes
itself from everyday computer use by ordinary users.
First, system software (e.g., servers, operating systems) is usually significantly larger and more complex than end-user application software.
For example, many of such software do not function independently but have sophisticated interactions and dependencies with co-running systems and
underlying infrastructure. However, unlike the use of application software, system configuration
requires administrators to have understanding of the system and its offered configurability (e.g.,
the impact of the settings).
On the other hand, system software is clearly not designed with novices in mind~\cite{velasquez:08}. Most configuration interfaces
ask administrators to directly set configuration parameters without helping them understand how these parameters
are used by the system and the potential impact of their settings. User manuals are supposed to help in this case. But the separation from
interfaces and manuals causes cognitive barriers due to the context switches,
not to mention that manuals are found to be mostly lengthy, and sometimes incomplete and obsolete~\cite{rabkin:11}.





Previous HCI studies treat system configuration as a type of end-user programming because administrators need
to edit files and write scripts to glue systems together~\cite{ko:11}. In fact, configuration and
programming have fundamental difference.
Unlike developers who have the opportunity to understand the internal organization of a system, administrators
usually view systems as black boxes without developing insights on how the systems are
designed and implemented. This significantly impairs their understanding of the systems
under configuration. Moreover, to diagnose a configuration problem, administrators cannot use interactive debugging tools (e.g., GDB) to check internal system states, 
but have to rely on
external manifestation (e.g., error code, system logs) to infer the root causes inside their settings.


Our goal in this paper is to provide an HCI view of today's configuration problems.
We articulate system configuration as a new HCI problem.
Moreover, we present the top obstacles to
correctly and efficiently configuring software systems, and more importantly their implications
on the design and implementation of new-generation configuration interfaces.
We do not discuss the configuration problems from a system perspective, including
testing misconfiguration vulnerabilities,
diagnosing misconfigurations from system failures,
tolerating configuration errors, and recovering from failures.
Our view is shaped in part by working in conducting research to defend systems against configuration errors since 2011
and in part by studying users' cognitive difficulties in configuring system software since September 2013.
Table~\ref{tab:summary} summarizes our hunches and findings on configuration problems and their implications for configuration interface design.

\section{2. WHAT IS CONFIGURATION AND WHAT IS NOT?}

\vspace{3pt}
\subsection{2.1 Definition of Configuration}

We define system configuration as the process of setting or tuning system parameters 
with the goal of converting an unsatisfying system behavior to a satisfying behavior.
The system behavior can be measured in many aspects, 
some of which are functionality, performance, security, reliability, diagnosability, etc.
Modern software systems often expose a wide range
of configuration parameters. For example, a typical Windows machine has more than 198,000
configurable Registry entries~\cite{wang:04}; Oracle 10g DBMS has 220 
initialization parameters and 1,477 tables of system parameters~\cite{keller:08}. By setting these
configuration parameters, administrators are able to control different aspects of system
behavior. 

Please note that configuration is a subset of administration operations. Other operations
include hardware management (e.g., plugging cables~\cite{jiang:08}), planning and provisioning (e.g., migrating
databases to new disks~\cite{barrett:04}), scripting and programming (e.g., writing scripts to
automate backup and monitoring jobs~\cite{haber:07}).
Configuration problem is particularly important among these operations because
configuration errors are reported as the largest category of operator errors~\cite{oppenheimer:03, nagaraja:04}.
In the study on three large Internet services~\cite{oppenheimer:03},
``{\it more than 50\% (and in one case nearly 100\%) of the operator errors that led to service
failures were configuration errors.}''

\subsection{2.2 Configuration Is Different from Programming}

Most previous studies on system administration and operations focus on the programming
perspective\footnote{Scripting is treated as one type of programming.}~\cite{barrett:04, haber:07, velasquez:08, ko:11}.
However, configuration is fundamentally different from programming or scripting.
In general, system administrators who perform configuration tasks do not have the same level of {\it view},
{\it understanding}, or {\it control} as the programmers who develop the systems.
This is reflected in two main aspects. First, unlike programmers, system administrators
do not write the code and usually do not (or cannot) read the code. Thus, it is hard
for them to exactly reason out the configuration requirements and the impact of their settings.
Documentation (e.g., user manuals) is supposed to help close this gap. Unfortunately, today's manuals are often
disappointing~\cite{rabkin:11}, probably because developers are not willing to spend effort writing manuals.
In addition, users may not be willing to read manuals line by line, especially
given their length (e.g., the user manual of MySQL-5.5 is 4502 pages long).
Second, when the configured system does not work as intended, administrators can hardly
debug the problems by themselves, especially for commercial systems where users do not have access
to source-code information. The lack of control makes the common debugging practices (e.g., inter-
actively examining program internal states) not applicable to misconfiguration troubleshooting.

\begin{figure}[tb!]
\centering
\includegraphics[width=0.4 \textwidth]{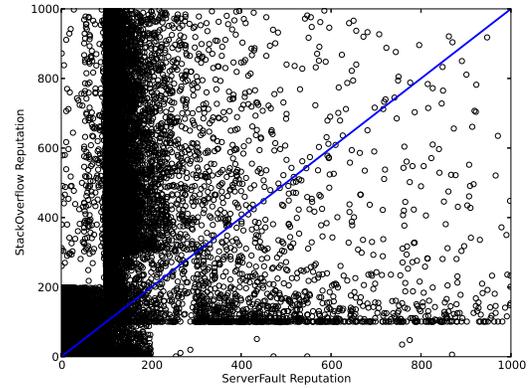}
\caption{{\bf System administration and programming are anti-correlated.} Each dot is a user's reputation score on ServerFault (x-axis) vs. on StackOverflow (y-axis).}
\label{fig:anti}
\end{figure}

Figure~\ref{fig:anti} plots users' reputation scores on ServerFault.com (SF) versus on StackOverflow.com (SO).
SF and SO are large Q \& A sites for ``{\it professional system and network administrators}'' and
``{\it professional and enthusiast programmers,}'' respectively. Both of the two sites are part of
the StackExchange Network from the same vendor. Thus, we can compare the information of a registered user
on the two sites. Note that the two sites follow the same reputation policy so that they are comparable.
Generally, the reputation score reflects the expertise of the user in the community: A higher score indicates
that the user is more active and capable. Figure~\ref{fig:anti} illustrates that system administration and
programming are anti-correlated. Most high-reputation users on SF has low reputation score on SO, and vice versa.
This indicates that system administrators and programmers form different communities, and their skill sets are different.
Thus, the previous studies and tools~\cite{ko:04, Ko:05, ko2:04, brandt:09, ko:11} designed for programmers may not be directly applicable to system administrators.

\section{3. SYSTEM CONFIGURATION AS A NEW HCI PROBLEM}

We argue that system configuration becomes a new HCI problem due to the shift
of the system administrator group. Consequently, previous studies, tools, and design principles are
not sufficient to solve todays' configuration problems.

\subsection{3.1 Defending Systems against Errors Is Not Enough}

As configuration errors are one of the major causes of system failures and anomalies,
the system community have worked on hardening systems against configuration errors for decades.
A rich set of tools and system mechanisms have been built to {\it fix} misconfiguration
vulnerabilities~\cite{keller:08, xu:13}, to {\it detect} certain types of configuration errors~\cite{feamster:05,zhang:14}, to {\it diagnose} system failures and anomalies
caused by configuration errors~\cite{attariyan:10, attariyan:12, wang:04, whitaker:04},
and to {\it tolerate} and {\it recover} from configuration errors~\cite{patterson:02}.
These tools have significantly improved the system defense to configuration errors.

We argue that these tools are not sufficient to solve today's configuration problems,
because they only deal with errors (and most of them cannot fix errors).
According to our study, configuration errors are a subset of configuration problems.
In many cases, system administrators fails to work out a solution and does not start on
the configuration process rather than committing errors.
Table~\ref{tab:activity} shows
the distribution of the activities during configuration where administrators encounter
problems and ask questions on ServerFault. 
The number is from 200 randomly sampled questions. 
We can see that problems related to errors only contribute
to 33.2\%. In our early study on configuration errors of a commercial storage system,
we randomly sampled 1000 customer cases~\cite{yin:11} and find that ``{\it more than half of them are
simply customer questions related to how the system should be configured.}''
Such cases are pruned out in the previous system studies because they are not considered as system problems.

Complementary to the system view, an HCI view of configuration problems is desired. 
To help administrators, we need to understand their cognitive problems of system configuration ---``{Which configuration tasks are found to be difficult and why so difficult?}''
``{Which configuration interface design is error-prone and causes frequent mistakes?}''
We believe that the fundamental solution to configuration problems should help system administrators
understand the systems and configuration parameters, and guide them towards correct configuration settings.



\begin{table}[tb!]
\centering 
\small
\begin{tabular}{ l | c } 
\hline
{\bf Configuration activity}   &      {\bf $\ \ $Percentage (\#)}$\ \ \ $ \\
\hline
Select software                &      22.3\% (35)            \\
Read manuals/tutorials         &      2.5\% (4)             \\
Find solutions                 &      38.9\% (61)            \\
\cellcolor{LightGray}{Fix active errors}       &  \cellcolor{LightGray}{23.6\% (37)}            \\
\cellcolor{LightGray}{Diagnose latent errors}  &  \cellcolor{LightGray}{9.6\% (15)}           \\
Validation                     &      3.2\% (5)            \\
\hline
\end{tabular}
\caption{{\bf Configuration activities on which users were stuck and asked questions on ServerFault}} 
\label{tab:activity} 
\end{table}



\subsection{3.2 Classic Interface Design Principles Are Not Sufficient}

Though the HCI community provides a rich set of principles of
UI design~\cite{nielsen:90, nielsen:94, norman:83a, norman:83b},
there is little understanding of configuration interface design.
The traditional UI design principles are mainly for end users who do not
need to deal with the system and implementation complexity.
Thus, some of the design principles might not be suitable
for configuration interfaces.
For example, one of Nielsen's 10 heuristics for UI design~\cite{nielsen:95} is
``{\it User Control and Freedom,}'' which is not a problem in
current file-based configuration interfaces (the problem instead is
that there are too many controllers and knobs).
In addition, some principles are too general and not specific to configuration
problems. For example, ``{\it Consistency and Standards}''~\cite{nielsen:95} is one
primary principle for interface design. But we do not have a good understanding of
{\it consistency} in the context of configuration, and which kind of {\it inconsistency} causes
administrators' cognitive errors. Thus, configuration-specific interface
design principles is desired.

We note that it is equally important to educate software developers
the importance of the configuration problems and the difficulties administrators
face when configuring the systems. Many developers still hold the opinion that system
administrators have sufficient knowledge of the system (because it is their jobs)
and work exactly as they expect. For example, a developer responded to
our report on a misconfiguration vulnerability~\footnote{Here ``vulnerabilities''
refers to bad system reactions to misconfigurations, such as crashes, hangs,
silent failures~\cite{xu:13}.} as follows~\cite{xu:13},

{\it ``If you work exactly and carefully, it does not matter;
if not, you should not maintain the server at all.''}

In the early 2000's, IBM researchers conducted a series of
ethnography field studies on professional system administrators and brought
insights on administration tools and
practices~\cite{maglio:03, barrett:04, kandogan:05, haber:07}.
Their focus was administration activities instead of configurations.
Most of the results are based on observing professional system administrators
from large enterprises and organizations,
which varies from many of today's administrators (e.g., from small business).
For example, all the system administrators they observed were parts of larger teams and spending
significant time (90\% of their time) using
telephone, instant messages, and emails communicating with their coworkers. For critical
operations, junior administrators worked side-by-side with experienced administrators.
However, this is not affordable to many non-professional administrators.
Moreover, most of the observed enterprises have comprehensive testing
infrastructure where administrators spent as much as a week testing all operations on a
series of test systems. This again is not affordable to small businesses.
Besides the environments and processes, many system administrators today do not come with a system's
or even computer science background. The following quotes was a part of a debate which happened between
an administrator and a developer of a open-source server software.

``{\it You are assuming that those who read that, understood what the context
of `user' was - I most assuredly did not until now. Unfortunately, many
of us don't come from UNIX backgrounds and though pick up on many
things, some things which seem basic to you guys elude us for some time.}''

Thus, many observations and conclusions of the early studies on professional administrators~\cite{hrebec:01,velasquez:08} might not hold on today's non- and semi-professional administrators.

{\bf {[I plan to conduct an online survey (e.g., on ServerFault)
to support these claims, including administrators' educational background,
profession, expertise, etc.]}}

\section{4. TOP CONFIGURATION OBSTACLES AND THEIR DESIGN IMPLICATIONS ON CONFIGURATION INTERFACES}

In this section, we offer a ranked list of obstacles for administrators to the correct system configuration.
Each obstacle is paired with an opportunity ---our thoughts on how to overcome the obstacle, ranging from
straightforward interface development to major research projects.

\begin{figure*}[tb!]
\centering
\includegraphics[width= 0.95\textwidth]{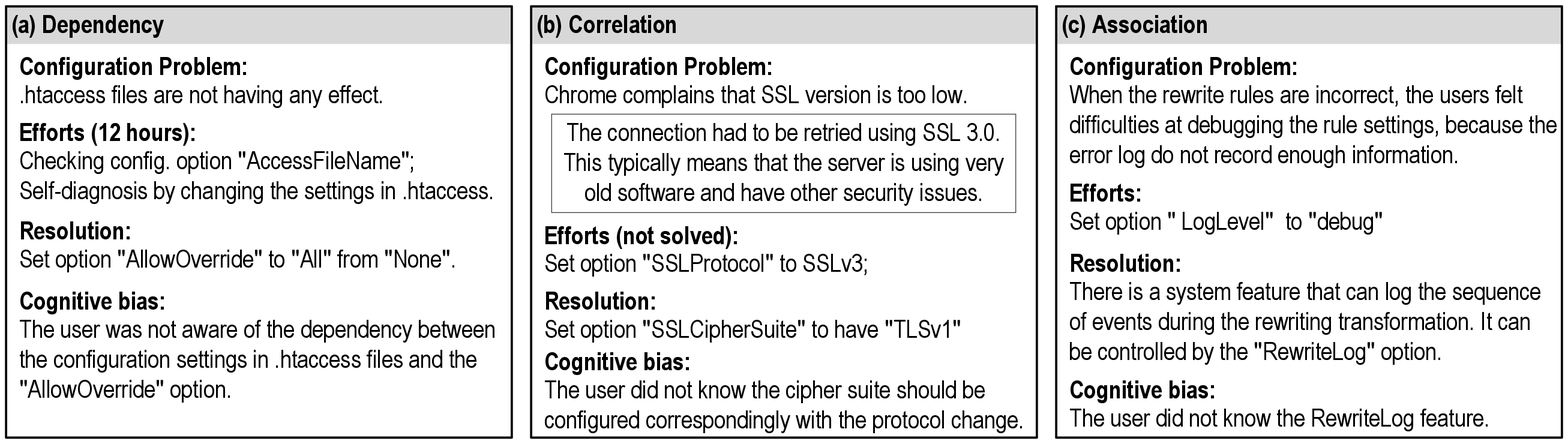}
\vspace{-5pt}
\caption{{\bf Lack of Guidance and Information.} Three types of information missed by administrators, resulting in their configuration problems}
\label{fig:knowledge}
\end{figure*}

\subsection{Obstacle 1: Lack of Guidance and Information}

Administrators configure their systems with intentions (i.e., expected system behavior), for example,
enabling certain functionalities, improving performance, enhancing security, etc.
To successfully perform configuration tasks, administrators have to address the following
two challenges:
\vspace{-8pt}
\begin{enumerate}
\item Finding the parameters related to the expected system behavior from hundreds of available parameters
\vspace{-3pt}
\item Setting correct values to the parameters with which the system would behave as expected.
\end{enumerate}
\vspace{-8pt}
Addressing either of them requires administrators to have a good understanding of
the provided configuration parameters and their impact on the system (which could be subtle).

The underlying assumption of the {\it de facto} file-based configuration interfaces
is that the administrators know exactly what parameters to set and how to set the parameter's values.
They do not help administrators address the above two challenges, but directly ask them to
edit the corresponding entries in the configuration file.
The responsibility of educating administrators and
helping them understanding systems is pushed to user manuals.
We observe that the separation of the interface for understanding the system (manual) and the interface for 
manipulating the system (text file)
cause cognitive difficulties and errors.

Figure~\ref{fig:knowledge} shows three real-world configuration problems in our
studied cases~\cite{xp:13}. All the three problems are failures in figuring out
the right configuration parameters from the entire parameter set. In fact, the administrators did
work on related parameters, and the resolutions were indeed documented in the user manuals
but somehow ignored by the administrators. The recent user survey of OS configuration~\cite{hubaux:12}
reports that the users' difficulties in activating the inactive parameters and 
determining the necessary parameters.

\begin{figure*}[tb!]
\centering
\includegraphics[width= 0.95\textwidth]{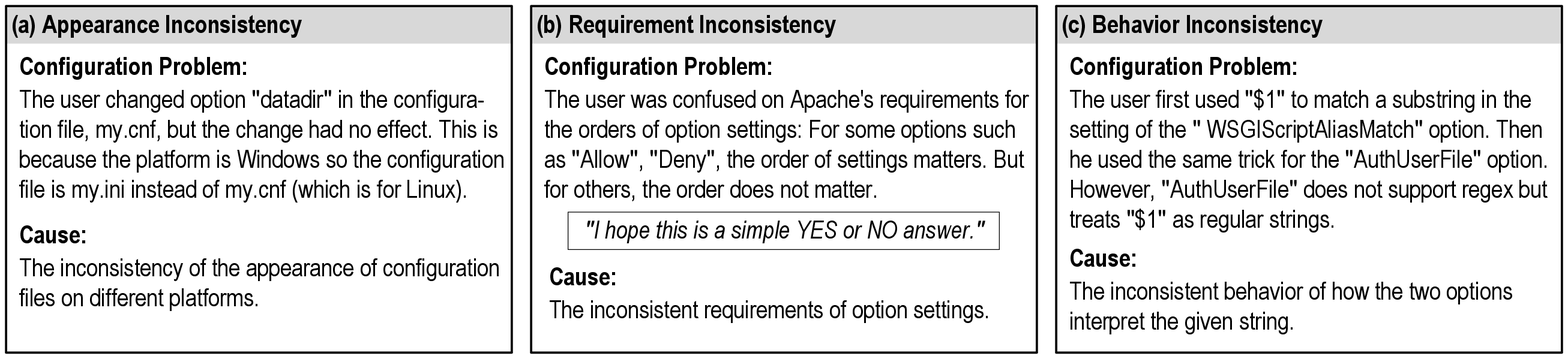}
\vspace{-5pt}
\caption{{\bf Inconsistency and Ambiguity.} Three types of inconsistencies that cause administrators' configuration problems}
\label{fig:inconsistency}
\vspace{3pt}
\end{figure*}



We advocate that configuration interfaces should take
the responsibility to guide administrators to address the two challenges
in the ``recognition-rather-than-recall'' manner~\cite{nielsen:90}.
To address the first challenge, i.e., finding needles in the haystack,
configuration interfaces should provide information that links administrators'
intention to the related configuration parameters.
We propose configuration interfaces to provide administrators with
the following three types of information during their configuration to help them
address the first challenge.
Figure~\ref{fig:knowledge} gives the examples of these three types of information.


\begin{itemize}
\item {\bf Dependency.} The dependency between different configuration
parameter and settings, including {\it control dependency} (``Parameter A has effects only when
parameter B is enabled.'') and {\it value dependency} (``Parameter A's setting should be less than parameter B's.'').

\item {\bf Correlation.} When an administrator sets
a configuration parameter, the correlated parameter (i.e., parameters should be or always be set together) should be listed.

\item {\bf Association.} The interface should remind the administrator of parameters that have similar
intention as the one set by her. For example, when the administrator is manipulating a security-related
parameter, it would be helpful to show her other parameters related to security.
\end{itemize}

To address the second challenge, the configuration interfaces should help administrators
set correct values which lead to the intended behavior. The following three types of information
should be provided by the configuration interface.

\begin{itemize}
\item {\bf Constraints.} They define the correctness requirements the parameter's setting should follow, for example,
                         data types, data unit, case sensitivity, formatting rules, etc. There are tools that can be leveraged to
                         extract configuration constraints~\cite{kiciman:04, xu:13, nadi:14} from source code
                         and user data.
\item {\bf Impact.} The interface should help administrators be aware of the potential system impact of the parameter settings, especially
                    the side effects. For example, enabling a security feature might degrade the performance due to extra computation overhead.
\item {\bf Examples.}  For complicated parameters such as rules and policies (e.g., regular expressions), providing working examples
                       would be very helpful to administrators, as shown by example-centric development~\cite{hartmann:07, brandt:10}.
\end{itemize}

We note that the lack of guidance and information in interface support is not only the obstacle for
end-user administrators but also for professional administrators (though may be less).
In the previous study on storage systems~\cite{jiang:09},
{\it ``a significant percentage of customer problems (11\%) are because customers lack sufficient knowledge about the system, which leads to misconfiguring the operating environment.''}


\subsection{Obstacle 2: Inconsistency and Ambiguity}

As widely accepted interface design principles~\cite{nielsen:90,norman:83b}, consistency and standard are surprisingly not
well maintained in today's configuration interfaces.
As a consequence, when administrators derive configuration settings by analogy with other similar parameters,
the ``derivation'' causes configuration problems (e.g., errors in the type of rule breakdowns~\cite{reason:90}) without the administrator realizing it.
In our study of Apache configuration problems posted on ServerFault.com~\cite{xp:13},
we observe that a significant portion ($\sim$30.2\%) were caused by inconsistency and ambiguity of the interfaces.

We observe that users' configuration problems caused by the inconsistency
of the following three aspects. This indicates that their consistency should
be evaluated and ensured with priority. Figure~\ref{fig:inconsistency} gives examples of real-world configuration problems
caused by the three types of inconsistency.

\begin{itemize}
\item {\bf Appearance.} It refers to users' perceived external appearance of the interface such as
parameter naming, configuration data formats, and control methods.

\item {\bf Requirement.} The requirements of configuration settings should be consistent, for example,
the granularity of parameter settings, case sensitivity, orders of commands.

\item {\bf Behavior.} The program behavior related to configuration should be consistent, for example,
the translation of user inputs, the reactions to configuration errors.
\end{itemize}

Inconsistency is mainly introduced when different sources contribute to the same code base.
Usually, configuration parameters from different system components are developed by different teams who might have different
habit and philosophy regarding configuration design and handling.
Unfortunately, we do not have standard software architecture nor design pattern to enforce the consistency
of configuration interface, but mainly rely on developers' preference.
Tooling support is desired to ensure consistency of the system on the whole.


\begin{figure*}[tb!]
\centering
\includegraphics[width= 0.95\textwidth]{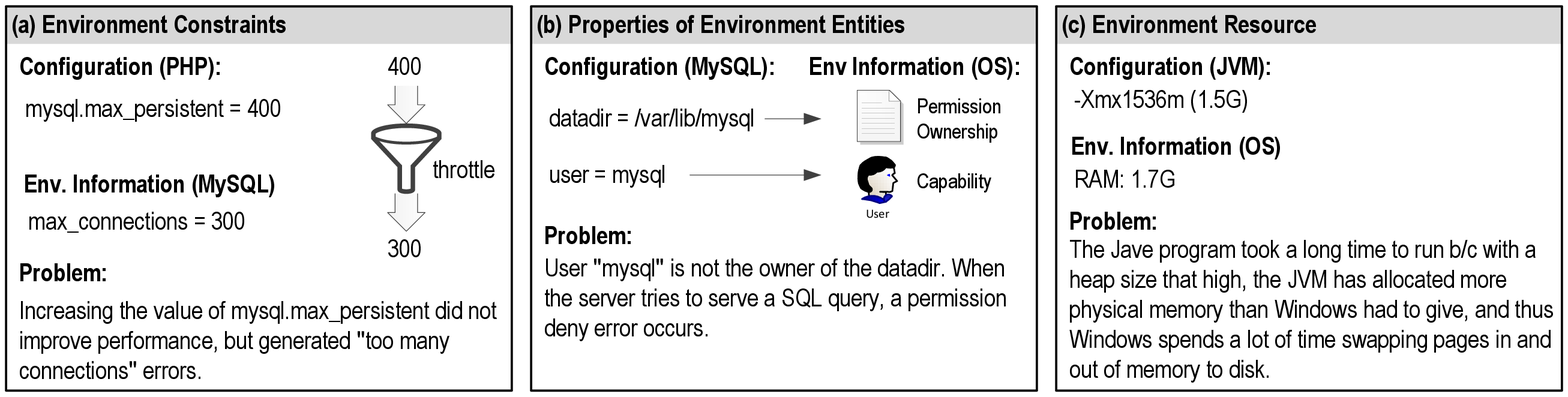}
\vspace{-5pt}
\caption{{\bf Lack of Environment Awareness.} Real-world examples of configuration problems caused by unawareness of correlated environment information.}
\label{fig:env}
\end{figure*}

\subsection{Obstacle 3: Complexity: Beyond The Capability}

Configuring software systems is complicated by nature due to the complexity of the systems under configuration
and the requirements of understanding them. Such complication causes obstacles to many of the end-user administrators
who is incapable (or have no time) to deal with such complexity but need the services.

We advocate system software to separate their configuration interfaces for different types of users
with different skill levels. We see attempts in the same vein for user manuals. For example,
IDG Books Inc. published two books about Apache server configuration in a row, for
``dummies''~\cite{coar:98} and professional administrators~\cite{kabir:99}, respectively.
The former provides tips to shortcut, warns potential problems, and highlights advanced
technical contents that can be skipped.
The latter serves as an ``ultimate shop manual'' with ``no introductory information,''
``no screen-shots,'' and ``no sidebars.''
Similar efforts should be made for the actual configuration interfaces.

In fact, configuration parameters differ by the levels of necessity. 
Some parameters are {\it must-to-set} for system functionalities. For example, to start a MySQL database
server, parameter ``datadir'' {\it must} be configured to specify the location of the data store on the file system.
On the other hand, most parameters are {\it nice-to-set} (e.g., for performance and security tuning).
Our hunch is that end-user administrators care less about performance and security, compared to professional administrators.
Thus, we think the two types of parameters should be treated differently.

Configuration parameters might also have different levels of understanding requirements.
For example, eight numeric configuration
parameters in Squid Web proxy server have the following explanation in their manual entries~\cite{xu:13}:

{\it ``Heavy voodoo here. I can't even believe you
are reading this. Are you crazy? Don't even
think about adjusting these unless you understand the algorithms in comm select.c first!''}

Such parameters should be excluded from the ``dummy'' interface. One common practice to simplify configuration
is to use metaphors with predefined templates. For example, Windows replaced rule-based firewall configuration
by an enumerative parameter with three values {\it ``home,''} {\it ``work,''} and {\it ``public.''} Such design significantly
simplified the firewall configuration~\cite{wool:04} while satisfying most daily needs.



Providing administrators with pre-configured templates for 
typical workload patterns is another way of simplifying the configuration tasks.
For example, MySQL provides template configuration files for {\it small}, {\it medium}, {\it large}, and 
{\it very large} systems. This saves the users' efforts of tuning MySQL performance according
to the hardware configuration.

\subsection{Obstacle 4: Lack of Environment Awareness}

One of the main challenges (and uniqueness) of system configuration is the
need to understand the system's running environments,
including the underlying software stacks,
network connection, co-running software, as well as their correlations.
Different from end-user applications, system software usually has multiple components
interacting with each other.
When the system fails to deliver the desired functionality, it is hard to
know which component/software is not configured properly.
In our previous study~\cite{yin:11}, we observe that a significant portion
(21.7\%$\sim$57.3\%) of configuration errors involve configurations beyond the
application itself or spanning across multiple hosts. Figure~\ref{fig:env}
shows three cross-component/-software real-world configuration problems.


These configuration problems are particularly hard for administrators to deal with.
In fact, none of them can be identified as configuration errors
without the awareness of correlated environment information.
However, the complexity of the system interaction with the environment causes the space of possible correlations
to be too large to permit an exhaustive exploration
(i.e., bounded rationality)~\cite{simon:56}.
Without tooling support, it is hard for human to always be
aware of the correlated information.

We advocate that configuration interfaces should help administrators be aware
of the execution environment by providing correlated environment information
during their configuration. The following three types of information is commonly
missed by administrators and causes their configuration problems. 
Thus, it should be provided by the interface. Figure~\ref{fig:env} gives three real-world examples of 
configuration problems caused by the missing of these information.

\begin{itemize}
\item {\bf Environment Constraints.} A configuration setting might be constrained by
the settings of co-running software (e.g., Figure~\ref{fig:env}) or underlying stack (e.g., OS limit).

\item {\bf Properties of Entities.} A configuration value might point to an entity in the execution environment (e.g., a file
in the file system). The interface should help users be aware of the properties of these entities.

\item {\bf Resource Information.} The configured resource allocation cannot exceed the available resource. 
\end{itemize}


Researchers have proposed methods (e.g.,~\cite{rama:09, zhang:14}) to obtain the environment information. 
Unfortunately, few of such information is used and integrated in configuration interfaces.




\subsection{Obstacle 5: Lack of Technical Support}

Many system administrators (especially those who manage open-source systems) use Internet
as the freely available technical support.
When they encounter configuration problems, they ask for support by posting their problems
on online Q \& A sites, user forums, and mailing lists.
Compared with commercial technical support services which designate
technical support engineers (TSEs) to resolve customers' cases,
Internet-based support services take advantage of the wisdom of the crowd.
The configuration problem is very likely to have been encountered and solved by other users.
In fact, with the diversity of hardware/software versions and the large space
of potential problems, it is difficult for TSEs to get familiar with all kinds of problems.
Internet-based support complements the knowledge limit.



\begin{figure}[tb!]
\centering
\includegraphics[width= 0.41\textwidth]{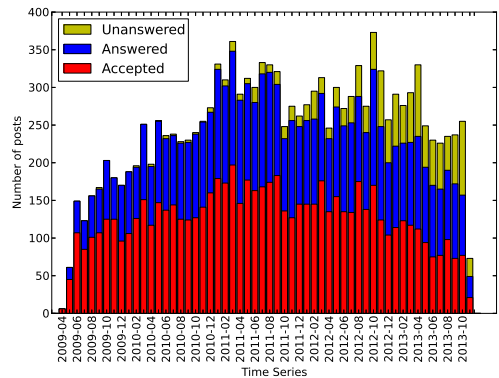}
\caption{\bf Number of unanswered, answered, and accepted posts on ServerFault in time series}
\label{fig:qa}
\end{figure}

On the other hand, today's Internet-based support has shortcomings.
First, many posts are never answered probably because of irresponsibility and neglect.
It is reported that many Q \& A sites have answer rates between 66\% and 90\%~\cite{dearman:10, mamykina:11}.
Though the success of StackOverflow shows that careful design of community
organization and user incentives can significantly increase answer rate and reduce answer time~\cite{mamykina:11},
the performance of Q \& A sites for configuration problems is still not satisfying.
Figure~\ref{fig:qa} shows the number of posts with accepted answers on ServerFault (using exactly the
same design as StackOverflow). We can see that only about 50\% posts have accepted answers.
Besides unanswered posts, there are many posts with answers not accepted.

\begin{figure}[tb!]
\centering
\includegraphics[width= 0.42\textwidth]{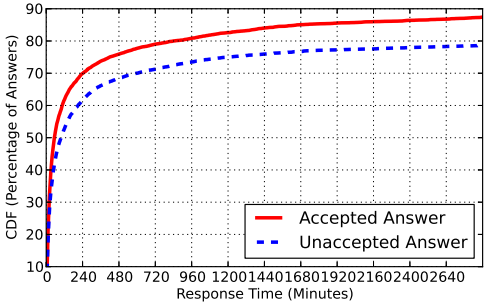}
\vspace{-5pt}
\caption{\bf Response time of accepted answers vs. unaccepted answers}
\label{fig:response_time}
\end{figure}


The usefulness of Internet-based technical support requires the answers to be timely and efficient.
Administrators usually have a time limit to solve configuration problems and look for solutions in this limited period.
Late support is not helpful even they can provide solutions. Figure~\ref{fig:response_time} shows the CDF (Cumulative Distribution Function)
of the response time of accepted answers and unaccepted answers\footnote{In this figure, we only select
answers for accepted posts. Unaccepted posts are still open, so the their response time might be biased.}. We can see the accepted answers have less response time.
In addition, a useful service should be efficient. If the first several answers do not pinpoint the root causes
or solutions of the problems, users might not be patient or confidence on the service. The cause of inefficiency
is that users fail to provide the information that is useful to diagnose the problem.
In some cases, other users keep asking the user to provide failure information, such as error logs, configuration
files, system state (e.g., from top, ifconfig), etc. An efficient, automatic, and privacy-preserving way of
collecting diagnosis information is desired (just as how commercial companies collect system information for technical support~\cite{jiang:09, lancaster:01, wang:03}).








\section{5. CONCLUSION}

The prevalence of today's configuration problems, as well as the severity
of the resulting configuration errors, reveals the importance of rethinking and redesigning
the configuration interfaces. The key shortcoming of today's configuration
interfaces is their inability of helping administrators understand the configuration
knobs (e.g., constraints, correlated environment information) 
and their impact on the system. With the evolved system
administrator group, it is desired to make configuration interfaces more
informative, instructive, user friendly, and concise.

In this paper, we summarize the top five cognitive obstacles that 
system administrators are facing towards correct and efficient system 
configuration. We believe that the next-generation configuration interfaces should be able
to help administrators overcome these obstacles.
Please note that the design implications of configuration interfaces discussed in the paper is orthogonal to
the form of the interface: They are applicable to GUI, command-line interface (CLI),
and even file-based interface. 
A good configuration interface does not mean a GUI wrapper around existing configuration files. 
Instead, a file-based interface 
can be designed informative and instructive by carefully organizing the structures and formats, 
adding comments, and highlighting errors in the file.
  

\bibliographystyle{abbrv}
{\footnotesize
\bibliography{ref}

\begin{thebibliography}{10}

\bibitem{andreessen:11}
M.~Andreessen.
\newblock {Why Software Is Eating The World}.
\newblock {\em Wall Street Journal}, August 2011.

\bibitem{attariyan:12}
M.~Attariyan, M.~Chow, and J.~Flinn.
\newblock {X-ray: Automating Root-Cause Diagnosis of Performance Anomalies in
  Production Software}.
\newblock In {\em Proceedings of the 10th USENIX Conference on Operating
  Systems Design and Implementation (OSDI'12)}, Hollywood, CA, USA, October
  2012.

\bibitem{attariyan:10}
M.~Attariyan and J.~Flinn.
\newblock {Automating Configuration Troubleshooting with Dynamic Information
  Flow Analysis}.
\newblock In {\em Proceedings of the 9th USENIX Conference on Operating Systems
  Design and Implementation (OSDI'10)}, Vancouver, BC, Canada, October 2010.

\bibitem{barrett:04}
R.~Barrett, E.~Kandogan, P.~P. Maglio, E.~Haber, L.~A. Takayama, and
  M.~Prabaker.
\newblock {Field Studies of Computer System Administrators: Analysis of System
  Management Tools and Practices}.
\newblock In {\em Proceedings of the 2004 ACM Conference on Computer Supported
  Cooperative Work (CSCW'04)}, Chicago, IL, USA, November 2004.

\bibitem{brandt:10}
J.~Brandt, M.~Dontcheva, M.~Weskamp, and S.~R. Klemmer.
\newblock {Example-Centric Programming: Integrating Web Search into the
  Development Environment}.
\newblock In {\em Proceedings of the 2011 ACM Conference on Human Factors in
  Computing Systems (CHI'10)}, Atlanta, GA, USA, April 2010.

\bibitem{brandt:09}
J.~Brandt, P.~J. Guo, J.~Lewenstein, M.~Dontcheva, and S.~R. Klemmer.
\newblock {Two Studies of Opportunistic Programming: Interleaving Web Foraging,
  Learning, and Writing Code}.
\newblock In {\em Proceedings of the 2009 ACM Conference on Human Factors in
  Computing Systems (CHI'09)}, Boston, MA, USA, April 2009.

\bibitem{se_domain}
{CircleID.com}.
\newblock {Misconfiguration Brings Down Entire .SE Domain in Sweden}.
\newblock 2009.
\newblock
  \\http://www.circleid.com/posts/misconfiguration\_brings\_down\_entire\\\_se\_domain\_in\_sweden/.

\bibitem{coar:98}
K.~A.~L. Coar.
\newblock {\em Apache Server For Dummies}.
\newblock IDG Books Worldwide Inc., 1998.

\bibitem{dearman:10}
D.~Dearman and K.~N. Truong.
\newblock {Why Users of Yahoo! Answers Do Not Answer Questions}.
\newblock In {\em Proceedings of the 2010 ACM Conference on Human Factors in
  Computing Systems (CHI'10)}, Atlanta, GA, USA, 2010.

\bibitem{feamster:05}
N.~Feamster and H.~Balakrishnan.
\newblock {Detecting BGP Configuration Faults with Static Analysis}.
\newblock In {\em Proceedings of the 2nd USENIX Symposium on Networked Systems
  Design and Implementation (NSDI'05)}, Boston, MA, USA, May 2005.

\bibitem{haber:07}
E.~M. Haber and J.~Bailey.
\newblock {Design Guidelines for System Administration Tools Developed through
  Ethnographic Field Study}.
\newblock In {\em Proceedings of the 2007 ACM Conference on Human Interfaces to
  the Management of Information Technology (CHIMIT'07)}, Cambridge, MA, USA,
  March 2007.

\bibitem{hartmann:07}
B.~Hartmann, L.~Wu, K.~Collins, and S.~R. Klemmer.
\newblock {Programming by a Sample: Rapidly Creating Web Applications with
  d.mix}.
\newblock In {\em Proceedings of the 20th Annual ACM Symposium on User
  Interface Software and Technology (UIST'07)}, Newport, Rhode Island, USA,
  October 2007.

\bibitem{hrebec:01}
D.~G. Hrebec and M.~Stiber.
\newblock {A Survey of System Administrator Mental Models and Situation
  Awareness}.
\newblock In {\em Proceedings of the 2001 Special Interest Group on Computer
  Personnel Research Annual Conference (SIGCPR'01)}, San Diego, CA, USA, 2001.

\bibitem{hubaux:12}
A.~Hubaux, Y.~Xiong, and K.~Czarnecki.
\newblock {A User Survey of Configuration Challenges in Linux and eCos}.
\newblock In {\em Proceedings of 6th International Workshop on Variability
  Modelling of Software-intensive Systems (VaMoS'12)}, Leipzig, Germany,
  January 2012.

\bibitem{jiang:09}
W.~Jiang, C.~Hu, S.~Pasupathy, A.~Kanevsky, Z.~Li, and Y.~Zhou.
\newblock {Understanding Customer Problem Troubleshooting from Storage System
  Logs}.
\newblock In {\em Proceedings of the 7th USENIX Conference on File and Storage
  Technologies (FAST'09)}, San Fransisco, CA, USA, February 2009.

\bibitem{jiang:08}
W.~Jiang, C.~Hu, Y.~Zhou, and A.~Kanevsky.
\newblock {Are Disks the Dominant Contributor for Storage Failures? A
  Comprehensive Study of Storage Subsystem Failure Characteristics}.
\newblock In {\em Proceedings of the 6th USENIX Conference on File and Storage
  Technologies (FAST'08)}, San Jose, CA, USA, February 2008.

\bibitem{kabir:99}
M.~J. Kabir.
\newblock {\em Apache Server Administrator's Handbook}.
\newblock IDG Books Worldwide Inc., 1999.

\bibitem{kandogan:05}
E.~Kandogan and E.~M. Haber.
\newblock {Security Administration Tools and Practices}, August 2005.

\bibitem{keller:08}
L.~Keller, P.~Upadhyaya, and G.~Candea.
\newblock {ConfErr: A Tool for Assessing Resilience to Human Configuration
  Errors}.
\newblock In {\em Proceedings of the 38th Annual IEEE/IFIP International
  Conference on Dependable Systems and Networks (DSN'08)}, Anchorage, Alaska,
  USA, June 2008.

\bibitem{kiciman:04}
E.~Kiciman and Y.-M. Wang.
\newblock {Discovering Correctness Constraints for Self-Management of System
  Configuration}.
\newblock In {\em Proceedings of the 1st International Conference on Autonomic
  Computing (ICAC'04)}, New York, NY, USA, May 2004.

\bibitem{ko:11}
A.~J. Ko, R.~Abraham, L.~Beckwith, A.~Blackwell, M.~Burnett, M.~Erwig,
  C.~Scaffidi, J.~Lawrance, H.~Lieberman, B.~Myers, M.~B. Rosson, G.~Rothermel,
  M.~Shaw, and S.~Wiedenbeck.
\newblock {The State of the Art in End-User Software Engineering}.
\newblock {\em ACM Computing Surveys}, 43(3):21--43, 2011.

\bibitem{ko:04}
A.~J. Ko and B.~A. Myers.
\newblock {Designing the Whyline: A Debugging Interface for Asking Questions
  about Program Behavior}.
\newblock In {\em Proceedings of the 2004 ACM Conference on Human Factors in
  Computing Systems (CHI'04)}, Vienna, Austria, April 2004.

\bibitem{Ko:05}
A.~J. Ko and B.~A. Myers.
\newblock {A framework and methodology for studying the causes of software
  errors in programming systems}.
\newblock {\em Journal of Visual Languages and Computing}, 16(1--2):41--84,
  2005.

\bibitem{ko2:04}
A.~J. Ko, B.~A. Myers, and H.~H. Aung.
\newblock {Six Learning Barriers in End-User Programming Systems}.
\newblock In {\em Proceedings of the 2004 IEEE Symposium on Visual Languages
  and Human-Centric Computing (VLHCC'04)}, Rome, Italy, September 2004.

\bibitem{lancaster:01}
L.~Lancaster and A.~Rowe.
\newblock {Measuring Real World Data Availability}.
\newblock In {\em Proceedings of the 15th Systems Administration Conference
  (LISA'01)}, San Diego, CA, USA, December 2001.

\bibitem{maglio:03}
P.~P. Maglio, E.~Kandogan, and E.~Haber.
\newblock {Distributed Cognition and Joint Activity in Collaborative Problem
  Solving}.
\newblock In {\em Proceedings of the 25th Annual Meeting of the Cognitive
  Science Society (CogSci'03)}, Boston, MA, USA, July 2003.

\bibitem{mamykina:11}
L.~Mamykina, B.~Manoim, M.~Mittal, G.~Hripcsak, and B.~Hartmann.
\newblock {Design Lessons from the Fastest Q \& A Site in the West}.
\newblock In {\em Proceedings of the 2011 ACM Conference on Human Factors in
  Computing Systems (CHI'11)}, Vancouver, BC, Canada, May 2011.

\bibitem{azure_outage}
R.~Miller.
\newblock {Microsoft: Misconfigured Network Device Caused Azure Outage}.
\newblock 2012.
\newblock
  \\http://www.datacenterknowledge.com/archives/2012/07/28/microsoft-misconfigured-network-device-caused-azure-outage/.

\bibitem{nadi:14}
S.~Nadi, T.~Berger, C.~K\"{a}stner, and K.~Czarnecki.
\newblock {Mining Configuration Constraints: Static Analyses and Empirical
  Results}.
\newblock In {\em Proceedings of the 36th International Conference on Software
  Engineering (ICSE'14)}, Hyderabad, India, 2014.

\bibitem{nagaraja:04}
K.~Nagaraja, F.~Oliveira, R.~Bianchini, R.~P. Martin, and T.~D. Nguyen.
\newblock {Understanding and Dealing with Operator Mistakes in Internet
  Services}.
\newblock In {\em Proceedings of the 6th USENIX Conference on Operating Systems
  Design and Implementation (OSDI'04)}, San Francisco, CA, USA, December 2004.

\bibitem{nielsen:94}
J.~Nielsen.
\newblock {Enhancing the Explanatory Power of Usability Heuristics}.
\newblock In {\em Proceedings of the ACM CHI 94 Human Factors in Computing
  Systems Conference (CHI'94)}, Boston, MA, USA, April 1994.

\bibitem{nielsen:95}
J.~Nielsen.
\newblock {10 Usability Heuristics for User Interface Design}.
\newblock \\http://www.nngroup.com/articles/ten-usability-heuristics/, 1995.

\bibitem{nielsen:90}
J.~Nielsen and R.~Molich.
\newblock {Heuristic Evaluation of User Interfaces}.
\newblock In {\em Proceedings of the ACM CHI 90 Human Factors in Computing
  Systems Conference (CHI'90)}, Seattle, WA, USA, April 1990.

\bibitem{norman:83b}
D.~A. Norman.
\newblock {Design Principles for Human-Computer Interfaces}.
\newblock In {\em Proceedings of the ACM CHI 83 Human Factors in Computing
  Systems Conference (CHI'83)}, Boston, MA, USA, December 1983.

\bibitem{norman:83a}
D.~A. Norman.
\newblock {Design Rules Based on Analyses of Human Error}.
\newblock {\em Communications of the ACM}, 26(4):254--258, April 1983.

\bibitem{oppenheimer:03}
D.~Oppenheimer, A.~Ganapathi, and D.~A. Patterson.
\newblock {Why Do Internet Services Fail, and What Can Be Done About It?}
\newblock In {\em Proceedings of the 4th USENIX Symposium on Internet
  Technologies and Systems (USITS'03)}, Seattle, WA, USA, March 2003.

\bibitem{patterson:02}
D.~Patterson, A.~Brown, P.~Broadwell, G.~Candea, M.~Chen, J.~Cutler,
  P.~Enriquez, A.~Fox, E.~Kiciman, M.~Merzbacher, D.~Oppenheimer, N.~Sastry,
  W.~Tetzlaff, J.~Traupman, and N.~Treuhaft.
\newblock {Recovery-Oriented Computing (ROC): Motivation, Definition,
  Techniques, and Case Studies}.
\newblock Technical Report UCB//CSD-02-1175, University of California Berkeley,
  March 2002.

\bibitem{rabkin:11}
A.~Rabkin and R.~Katz.
\newblock {Static Extraction of Program Configuration Options}.
\newblock In {\em Proceedings of the 33th International Conference on Software
  Engineering (ICSE'11)}, Waikiki, Honolulu, Hawaii, USA, May 2011.

\bibitem{rabkin:13}
A.~Rabkin and R.~Katz.
\newblock {How Hadoop Clusters Break}.
\newblock {\em IEEE Software Magazine}, 30(4):88--94, July 2013.

\bibitem{rama:09}
V.~Ramachandran, M.~Gupta, M.~Sethi, and S.~R. Chowdhury.
\newblock {Determining Configuration Parameter Dependencies via Analysis of
  Configuration Data from Multi-tiered Enterprise Applications}.
\newblock In {\em Proceedings of the 6th International Conference on Autonomic
  Computing and Communications (ICAC'09)}, Barcelona, Spain, June 2009.

\bibitem{reason:90}
J.~Reason.
\newblock {\em {Human Error}}.
\newblock Cambridge University Press, October 1990.

\bibitem{simon:56}
H.~Simon.
\newblock {Rational choice and the structure of the environment}.
\newblock {\em Psychological Review}, 63(2):129--138, March 1956.

\bibitem{amazon_cloud_crash}
K.~Thomas.
\newblock {Thanks, Amazon: The Cloud Crash Reveals Your Importance}.
\newblock 2011.
\newblock
  \\http://www.pcworld.com/article/226033/thanks\_\\amazon\_for\_making\_possible\_much\_of\_the\_internet.html.

\bibitem{velasquez:08}
N.~F. Velasquez, S.~Weisband, and A.~Durcikova.
\newblock {Designing Tools for System Administrators: An Empirical Test of the
  Integrated User Satisfaction Model}.
\newblock In {\em Proceedings of the 22nd Large Installation System
  Administration Conference (LISA'08)}, San Diego, CA, USA, November 2008.

\bibitem{wang:04}
H.~J. Wang, J.~C. Platt, Y.~Chen, R.~Zhang, and Y.-M. Wang.
\newblock {Automatic Misconfiguration Troubleshooting with PeerPressure}.
\newblock In {\em Proceedings of the 6th USENIX Conference on Operating Systems
  Design and Implementation (OSDI'04)}, San Francisco, CA, USA, December 2004.

\bibitem{wang:03}
Y.-M. Wang, C.~Verbowski, J.~Dunagan, Y.~Chen, H.~J. Wang, C.~Yuan, and
  Z.~Zhang.
\newblock {STRIDER: A Black-box, State-based Approach to Change and
  Configuration Management and Support}.
\newblock In {\em Proceedings of the 17th Large Installation Systems
  Administration Conference (LISA'03)}, San Diego, CA, USA, October 2003.

\bibitem{whitaker:04}
A.~Whitaker, R.~S. Cox, and S.~D. Gribble.
\newblock {Configuration Debugging as Search: Finding the Needle in the
  Haystack}.
\newblock In {\em Proceedings of the 6th USENIX Conference on Operating Systems
  Design and Implementation (OSDI'04)}, San Francisco, CA, USA, December 2004.

\bibitem{wool:04}
A.~Wool.
\newblock {A Quantitative Study of Firewall Configuration Errors}.
\newblock {\em IEEE Computer}, 37(6):62--67, June 2004.

\bibitem{xp:13}
T.~Xu and V.~Pandey.
\newblock {Why System Administrators Fail at Configuration?}
\newblock {\em HCI Course Project Report, UC San Diego}, December 2013.

\bibitem{xu:13}
T.~Xu, J.~Zhang, P.~Huang, J.~Zheng, T.~Sheng, D.~Yuan, Y.~Zhou, and
  S.~Pasupathy.
\newblock {Do Not Blame Users for Misconfigurations}.
\newblock In {\em Proceedings of the 24th Symposium on Operating System
  Principles (SOSP'13)}, Farmington, PA, USA, November 2013.

\bibitem{yin:11}
Z.~Yin, X.~Ma, J.~Zheng, Y.~Zhou, L.~N. Bairavasundaram, and S.~Pasupathy.
\newblock {An Empirical Study on Configuration Errors in Commercial and Open
  Source Systems}.
\newblock In {\em Proceedings of the 23rd ACM Symposium on Operating Systems
  Principles (SOSP'11)}, Cascais, Portugal, October 2011.

\bibitem{zhang:14}
J.~Zhang, L.~Renganarayana, X.~Zhang, N.~Ge, V.~Bala, T.~Xu, and Y.~Zhou.
\newblock {EnCore: Exploiting System Environment and Correlation Information
  for Misconfiguration Detection}.
\newblock In {\em Proceedings of the 19th International Conference on
  Architecture Support for Programming Languages and Operating Systems
  (ASPLOS'14)}, Salt Lake City, UT, USA, March 2014.

\end{thebibliography}
}
\end{document}